\documentclass{emulateapj}
\usepackage{natbib}
\usepackage{graphicx}
\usepackage{amsmath}
\usepackage{float}
\newcommand{\nustar} {\textit{NuSTAR}}

\shorttitle{Pulsation dropout in LMC X-4}
\shortauthors{Brumback et al.}
\slugcomment{Accepted for publication in The Astrophysical Journal Letters}

\begin{document}

\title{Discovery of pulsation dropout and turn-on during the high state of the accreting X-ray pulsar LMC X-4}

\author{McKinley C. Brumback{\altaffilmark{1}}, Ryan C. Hickox{\altaffilmark{1}}, Matteo Bachetti{\altaffilmark{2}}, Ralf Ballhausen{\altaffilmark{3}}, Felix S. F\"urst{\altaffilmark{4}}, Sean Pike{\altaffilmark{5}}, Katja Pottschmidt{\altaffilmark{6,7}}, John A. Tomsick{\altaffilmark{8}}, J\"orn Wilms{\altaffilmark{3}}  }

\altaffiltext{1} {Department of Physics \& Astronomy, Dartmouth College, 6127 Wilder Laboratory, Hanover, NH 03755, USA} 
\altaffiltext{2} {INAF/Osservatorio Astronomico di Cagliari, via della Scienza 5, I-09047 Selargius (CA), Italy 0000-0002-4576-9337}
\altaffiltext{3} {Dr. Karl Remeis-Sternwarte and Erlangen Centre for Astroparticle Physics, Sternwartstrasse 7, 96049 Bamberg, Germany} 
\altaffiltext{4} {European Space Astronomy Centre (ESA/ESAC), Operations Department, Villanueva de la Ca$\tilde{\text{n}}$ada Madrid, Spain} 
\altaffiltext{5} {Cahill Center for Astronomy and Astrophysics, California Institute of Technology, 1216 East California Boulevard, Pasadena, CA 91125, USA}
\altaffiltext{6} {CRESST, Department of Physics and Center for Space Science and Technology, UMBC, Baltimore, MD 210250, USA} 
\altaffiltext{7} {NASA Goddard Space Flight Center, Code 661, Greenbelt, MD 20771, USA} 
\altaffiltext{8} {Space Sciences Laboratory, University of California, Berkeley, 7 Gauss Way, Berkeley, CA 94720, USA}

\begin{abstract}
Two \nustar\ observations of the luminous X-ray pulsar LMC X-4 in October and November 2015 captured several bright accretion flares from this source, which has a long history of stable pulse and superorbital behavior. We present a timing analysis of these data in which we detect a rapid pulse ``turn-on" in association with the accretion flares, during which the source reaches super-Eddington luminosities. Pulsations, which are normally seen from this source, are found to only occur for approximately one hour before and during the bright flares. Beyond one hour before and after the flares, we find pulsations to be weak or nonexistent, with fractional RMS amplitudes of less than 0.05. At the onset of the flare, the pulse profiles exhibit a phase shift of 0.25 cycles that could be associated with a change in the emission geometry. This increase in pulse strength occurring well before the flare cannot be explained by the propeller effect, and potentially offers a connection between the magnetic properties of pulsars that accrete close to their Eddington limits and ULX pulsars.

\end{abstract}

\section{Introduction} \label{sec:int}

Accreting X-ray pulsars consist of a magnetized neutron star and a stellar companion, which donates mass to the compact object either through Roche lobe overflow or the formation of outflow winds or disks ({e.g.\ \citealt{nagase2001}). Within the neutron star's magnetosphere, the magnetic pressure dominates the ram pressure of the accretion disk, and therefore gas falls along magnetic field lines onto the poles of the neutron star while releasing X-rays (e.g.\ \citealt{romanova2004}). The characteristic pulsations arise from the misalignment between the rotation and magnetic axes. The magnetic field geometry, orientation of the pulsar, and accretion rate all contribute to the shape and strength of the pulsations, which are generally stable over the time scale of a science observation.  

Some pulsars exhibit changes in pulse intensity or profile on short time scales, and these changes are most commonly driven by changes in accretion rate. A decrease in accretion rate and thus flux can result in the non-detection of pulsations from the weakened pulsar beam (e.g.\ \citealt{bozzo2008}). One way to bring about this effect is the ``propeller regime" first described by \cite{illarionov1975}, in which the accretion pressure no longer overcomes the magnetic field pressure and a centrifugal barrier forms, which halts accretion. The onset of the propeller regime has been used to describe drastic drops in flux and loss of pulsations from a number of pulsar X-ray binaries including 4U 0115$+$63 and V 0332$+$53 (\citealt{tsygankov2016b}), SMC X-2 (\citealt{lutovinov2017}), and EXO 2030$+$375 (\citealt{fuerst2017}). 

A clumpy wind environment can also cause a decrease in accretion rate, as is seen in the high mass X-ray binary (HMXB) Vela X-1 (\citealt{quaintrell2003,kreykenbohm2008,fuerst2010}). This source exhibits strong X-ray variability, including off states during which pulsations are not detected (\citealt{staubert2004, kreykenbohm2008,fuerst2010}). Hydrodynamic modeling suggests that areas of low density within the stellar wind could shut off accretion, and therefore pulsations, on short time scales (\citealt{manousakis2015}).

Another cause of pulsation dropout is absorption. The low mass X-ray binary Her X-1 contains a warped, precessing inner accretion disk, which gives rise to its 35 d superorbital cycle (\citealt{giacconi1973, ramsay2002}). The pulse profile varies in both shape and amplitude during this cycle, with pulsations dropping out during the low states due to scattering of the pulsar beam through the occulting inner accretion disk (\citealt{deeter1998,kuster2001}). Yet increased absorption cannot always describe pulse dropout, as in the case of HMXB GX 301-2. \cite{fuerst2011} found two drops in flux that were accompanied by a stop in pulsations but with no evidence of absorption in the X-ray spectra.

Pulse dropout phenomena have recently also been observed in some ultra-luminous X-ray (ULX) pulsars: M82 X-2 (\citealt{bachetti2014}), NGC 7793 P13 (\citealt{fuerst2016b, israel2017a}), NGC 5907 ULX-1 (\citealt{israel2017b}), and NGC 300 ULX1 (\citealt{carpano2018}). These sources have luminosities on the order of $10^{40}$ erg s$^{-1}$, hundreds of times the Eddington luminosity for neutron stars. These sources may have strong magnetic fields which allow accretion to reach these extreme super-Eddington levels (e.g.\ \citealt{eksi2015,dallosso2015,mushtukov2015}). In M82 X-2 and NGC 5907 ULX-1, the pulsations are not detected across the entire observation (\citealt{bachetti2014,israel2017a}). Although the propeller effect has been proposed (\citealt{tsygankov2016a}), the mechanism for pulsation dropout is not yet known and could give important clues to the role of the magnetic field in super-Eddington accretion.

Here we present a timing analysis of the eclipsing HMXB LMC X-4, which consists of a $1.57 \pm 0.11$ M$_\sun$ neutron star orbiting a $18 \pm 1$ M$_\sun$ O8 III companion star with a 1.4 day period (errors are $1\sigma$)(\citealt{kelley1983,falanga2015}). LMC X-4 is a bright binary with a characteristic X-ray luminosity of $\sim$ 2 $\times 10^{38}$ erg s$^{-1}$, close to the Eddington limit for neutron stars (\citealt{moon2003}). The spin period of the pulsar is  $\approx$13.5 s and has been found to have near-periodic variations with an average period derivative of $1 \times 10^{-10}$ ss$^{-1}$ (\citealt{white1978, molkov2017}). Pulsations have been previously detected in both the high and low state of the 30 d superorbital cycle (e.g.\ \citealt{hung2010}). Like Her X-1, LMC X-4 exhibits a periodic change in luminosity on a 30 d time scale due to a warped inner accretion disk (\citealt{lang1981,hung2010}), and exhibits bright flares that can reach super-Eddington luminosities of $\sim$ $10^{39}$ erg s$^{-1}$ (\citealt{kelley1983,levine1991,levine2000,moon2001,moon2003}). These flares occur non-periodically, typically last $\sim$1000 s, and show softening of the X-ray spectrum (e.g.\ \citealt{levine2000}). Pulsations have been observed consistently during flaring events, with the pulse profile intensity increasing with source luminosity (\citealt{moon2003}). Changes in pulse phase have also been observed during some flaring events (\citealt{dennerl1989,beri2017}). This Letter focuses on the first known detection of pulsation dropout and turn-on in LMC X-4. A spectral analysis of these data has been presented by \cite{shtykovsky2018}. 

\section{Data Analysis and Results} \label{sec:results}

\subsection{Observation and data analysis} \label{sec:obs}

The data for this Letter consist of two observations with the Nuclear Spectroscopic Telescope Array (\nustar) taken on 30 October 2015 (ObsID 30102041002, hereafter Observation 02) and 27 November, 2015 (ObsID 30102041008, hereafter Observation 08). These observations have exposure times of 24.5 ks and 20.3 ks, respectively. \nustar\ consists of twin grazing incidence telescopes sensitive to 3-79 keV (\citealt{harrison2013}). These observations were taken 29 days apart during consecutive turn ons of LMC X-4's superorbital cycle at superorbital phase $\approx$0.8, using the \cite{molkov2015} superorbital ephemeris. We used the \cite{levine2000} ephemeris to find that Observation 02 spans the orbital phase range 0.3--0.6 and Observation 08 spans the range of 0.4--0.7, where orbital phase zero corresponds to the mid-eclipse time.  We reduced these data using NuSTARDAS version 1.8.0 (part of HEASoft version 6.22). Source spectra were extracted from a circular region with an 120\arcsec\ radius centered on the source. Background spectra were extracted from a circular region of the same radius located on the other side of the field of view. The barycentric correction was applied using the FTOOL {\fontfamily{qcr}\selectfont barycorr} (using the JPL DE-200 ephemeris), and the photon arrival times were further corrected for the orbit of LMC X-4 using the ephemeris defined by \cite{levine2000}.

These data were part of a series of four observations to monitor LMC X-4 at various superorbital phases. The other two observations in this series (ObsID 30102041004 and 30102041006) are excluded from this analysis because they do not show accretion flares or rapid changes in pulse behavior.

\subsection{Light curve and pulse profiles}
\label{sec:curve}

\begin{figure*}
\plottwo{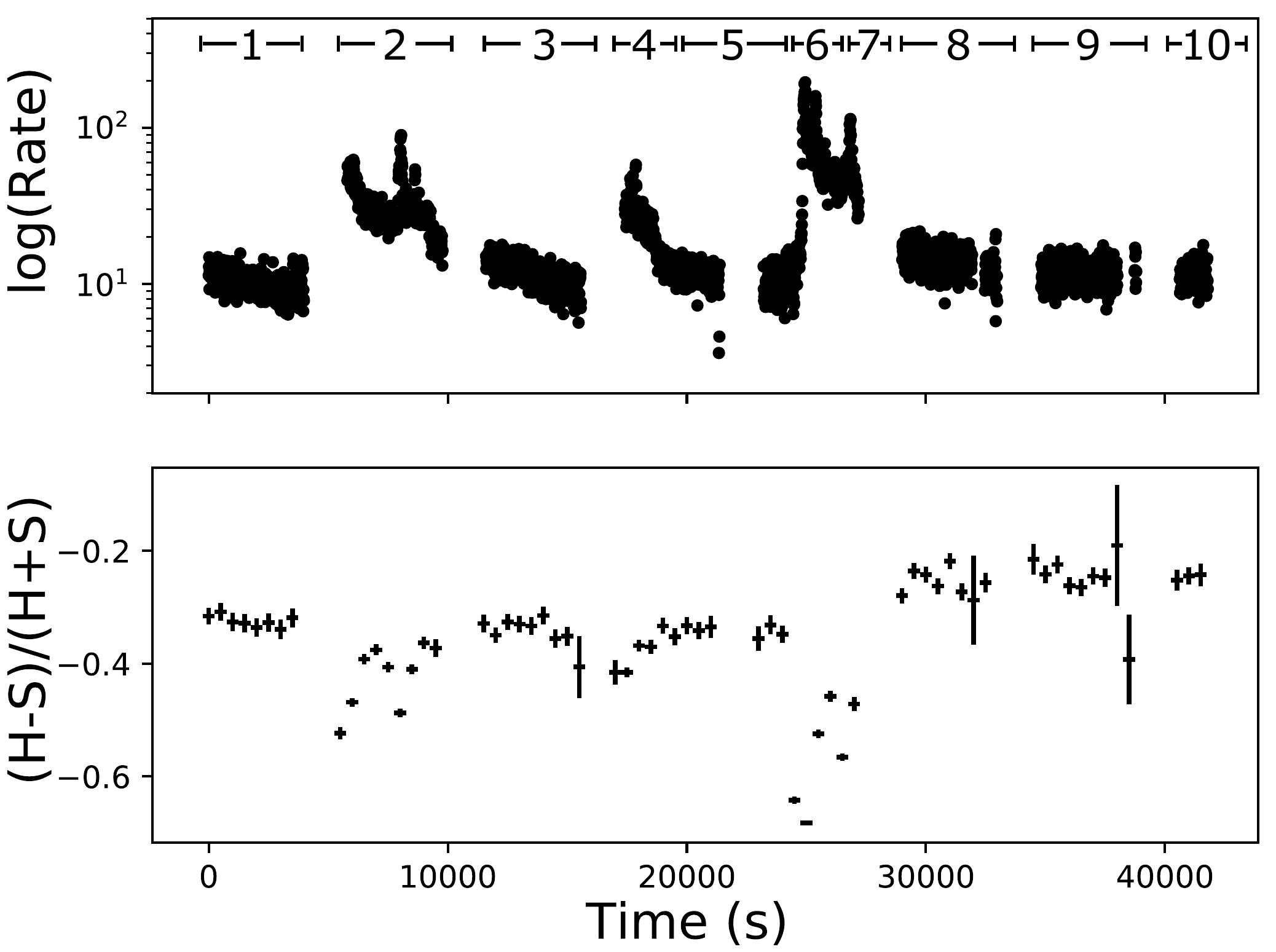}{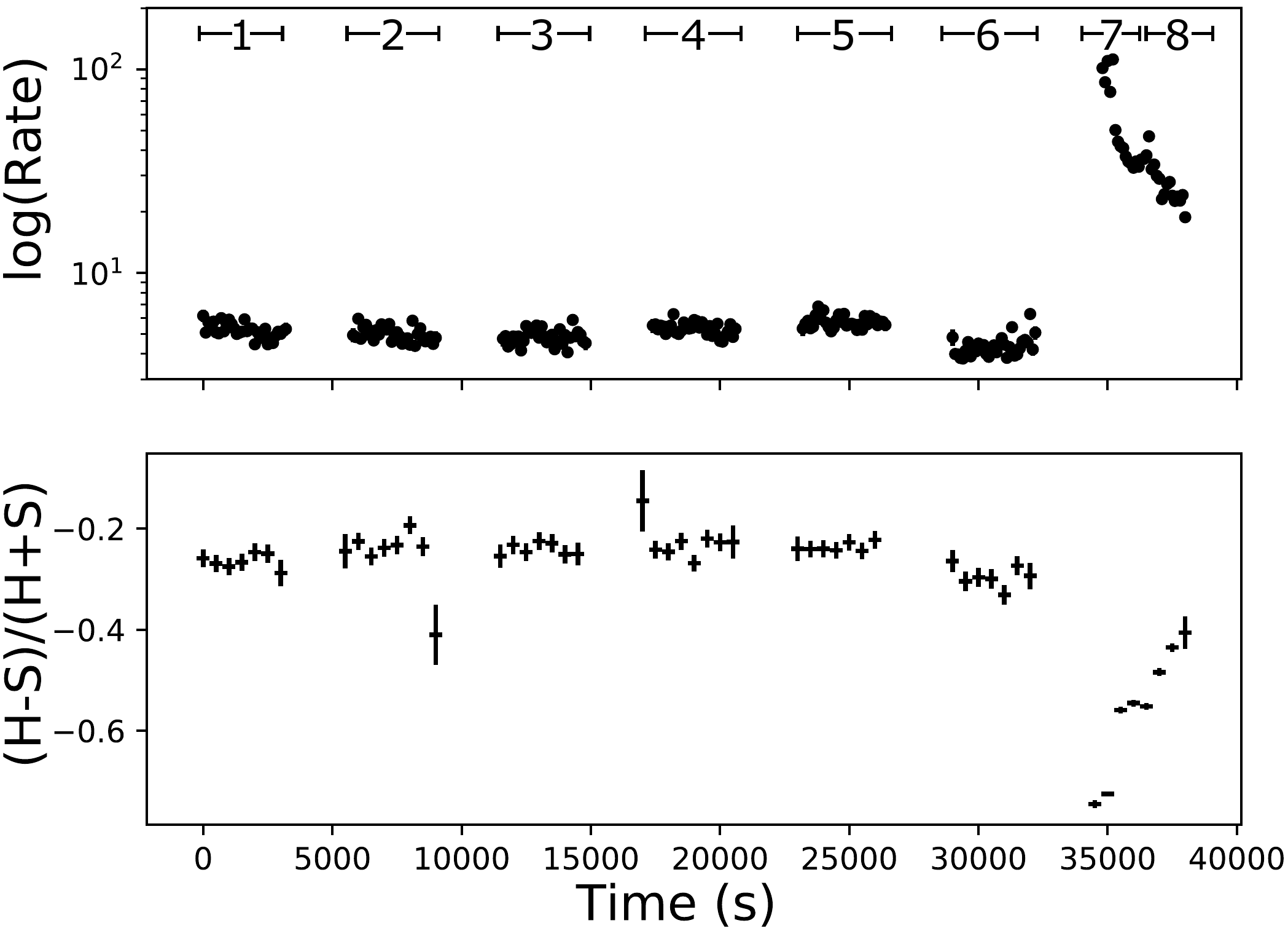}
\caption{Left: \nustar\ 3--79 keV light curve and hardness ratio for Observation 02 (30102041002), where $t=0$ corresponds to 2015 Oct 30, 01:01 UTC. Three large flares are present in the light curve. Right: \nustar\ 3--79 keV light curve and hardness ratio for Observation 08 (30102041008), where $t=0$ corresponds to 2015 Nov 27, 09:16 UTC. A large flare is present at the end of the light curve. The numbers reference time intervals of the light curve as discussed in Section \ref{sec:results}. The hardness ratios for both observations were calculated from 500 s bins, defined as in Section~\ref{sec:curve}. In both observations, the flares show significant spectral softening. Points with large error bars correspond to bins with shorter than 500 s exposure. }
\label{fig:lchr}
\end{figure*}

The 3--79 keV light curves of both observations are shown in the top panels of Figure \ref{fig:lchr}, where the numbered intervals mark time intervals of interest in this analysis. Three bright accretion flares are visible in the light curve of Observation 02, and one flare is visible at the end of Observation 08. In Observation 02, the average count rate in intervals 8, 9, and 10 is 12.8 counts s$^{-1}$. The intervals immediately preceding the flares (intervals 1, 3, and 5) have an average count rate of 10.1 counts s$^{-1}$, or approximately 20\% lower than the flux in intervals 8, 9, and 10. In Observation 08, intervals 1--5 have a mean count rate of 5.2 counts s$^{-1}$. Interval 6, preceding the flare shows an $\approx$20\% decrease in flux, with a mean count rate of 4.3 counts s$^{-1}$. The background light curve count rate in Observation 02 and Observation 08 are negligible, with a mean value of 0.06 counts s$^{-1}$ and 0.02 counts s$^{-1}$, respectively.

The flares show similar characteristics in both observations. In Observation 02, only the third flare has been captured at its maximum flux, which corresponds to a \nustar\ count rate of 195 counts s$^{-1}$. The flare in Observation 08 has a peak count rate of 112 counts s$^{-1}$. All flares observed in these two observations have a duration of $\approx$3000 s, and decay rapidly over a time of $\approx$1000 s. 

Assuming a distance of 50 kpc, we calculated the 3--60 keV luminosity of both observations. In Observation 02, the non-flared emission has L$_{\text{X}} \approx 3 \times 10^{38}$ erg s$^{-1}$, while the third and brightest flare reaches a luminosity of $\sim 1 \times 10^{39}$ erg s$^{-1}$. In Observation 08, the luminosity of the first 6 time intervals is $\sim 2 \times 10^{38}$ erg s$^{-1}$ and the luminosity of the flare is $\sim 1 \times 10^{39}$ erg s$^{-1}$.

The flares in both observations are significantly softer in spectral shape compared to the non-flared emission, as shown by the hardness ratios in Figure \ref{fig:lchr}. We calculated the hardness ratio as $(H-S)/(H+S)$ where $H$ and $S$ are the number of counts in the hard (10--20 keV) and soft (3--10 keV) bands, respectively. In Observation 02, the hardness ratio is generally constant in the intervals immediately preceding flares (intervals 1, 3, and 5). These intervals are approximately 20\% softer on average than the intervals following the final flare (intervals 8--10). A similar trend is seen in Observation 08, where the hardness ratio is generally constant across the first 5 time intervals, but interval 6, which precedes the flare, appears to be approximately 20\% softer on average than the previous intervals.

The pulse profiles presented in this analysis were created using the FTOOL {\fontfamily{qcr}\selectfont efold}, which folds the light curve on a given period. We required 20 bins per phase for these pulse profiles. We used an epoch folding technique to find the best period and a Monte Carlo simulation of 500 light curves to find the error. Due to complex pulse behavior in these observations (see Figures \ref{fig:pulseprofs4}, \ref{fig:pulseprofs1}, \ref{fig:pfvst}, and Section \ref{sec:pf}) we only folded the parts of each observation that are strongly pulsed. For Observation 02, we folded the first $\approx$28 ks to find a spin period of 13.5033 $\pm$ 0.0001 s (reference epoch 57325.3 MJD). For Observation 08, we folded the last $\approx$4000 s of the light curve to find our best period value of 13.5003 $\pm$ 0.0007 s (reference epoch 57353.8 MJD). These spin periods are used on their respective observations throughout this analysis. We note that this period is different from the value found by \cite{shtykovsky2018}, who used ObsIDs 30102041004 and 30102041006 (reference epoch 57334.4 MJD) to determine their pulse period of 13.50124 $\pm$ 0.00005 s. Our period is consistent with theirs at the 1.3$\sigma$ level, and within the long term period derivative found by \cite{molkov2017}.

\begin{figure*}
\centering
\includegraphics[scale=0.67]{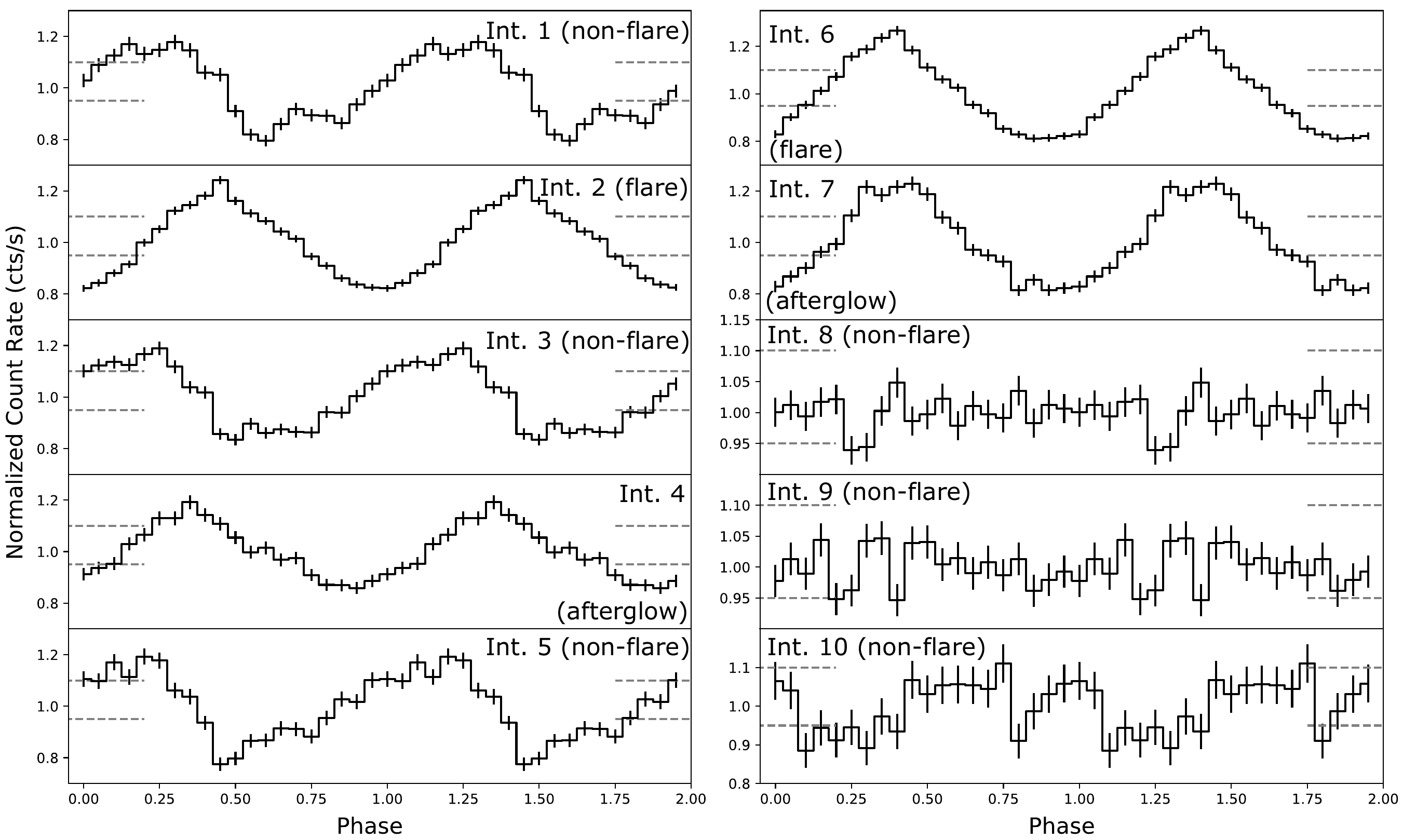}
\caption{Pulse profiles for Observation 02 in the full \nustar\ energy band, where the count rate has been normalized to the mean count rate in each interval. The horizontal dashed lines are plotted at the same rate in every panel to indicate the scale. The strength of the pulses dramatically drops off following the third and final flare. Changes in pulse shape are also evident, from a singly pointy peak in the flare and afterglow intervals (intervals 2, 4, 6, and 7), to more complex shapes in intervals 1, 3, and 5. A significant phase shift occurs between the intervals that contain flares and the intervals that do not.}
\label{fig:pulseprofs1}
\end{figure*}

\begin{figure*}
\centering
\includegraphics[scale=0.67]{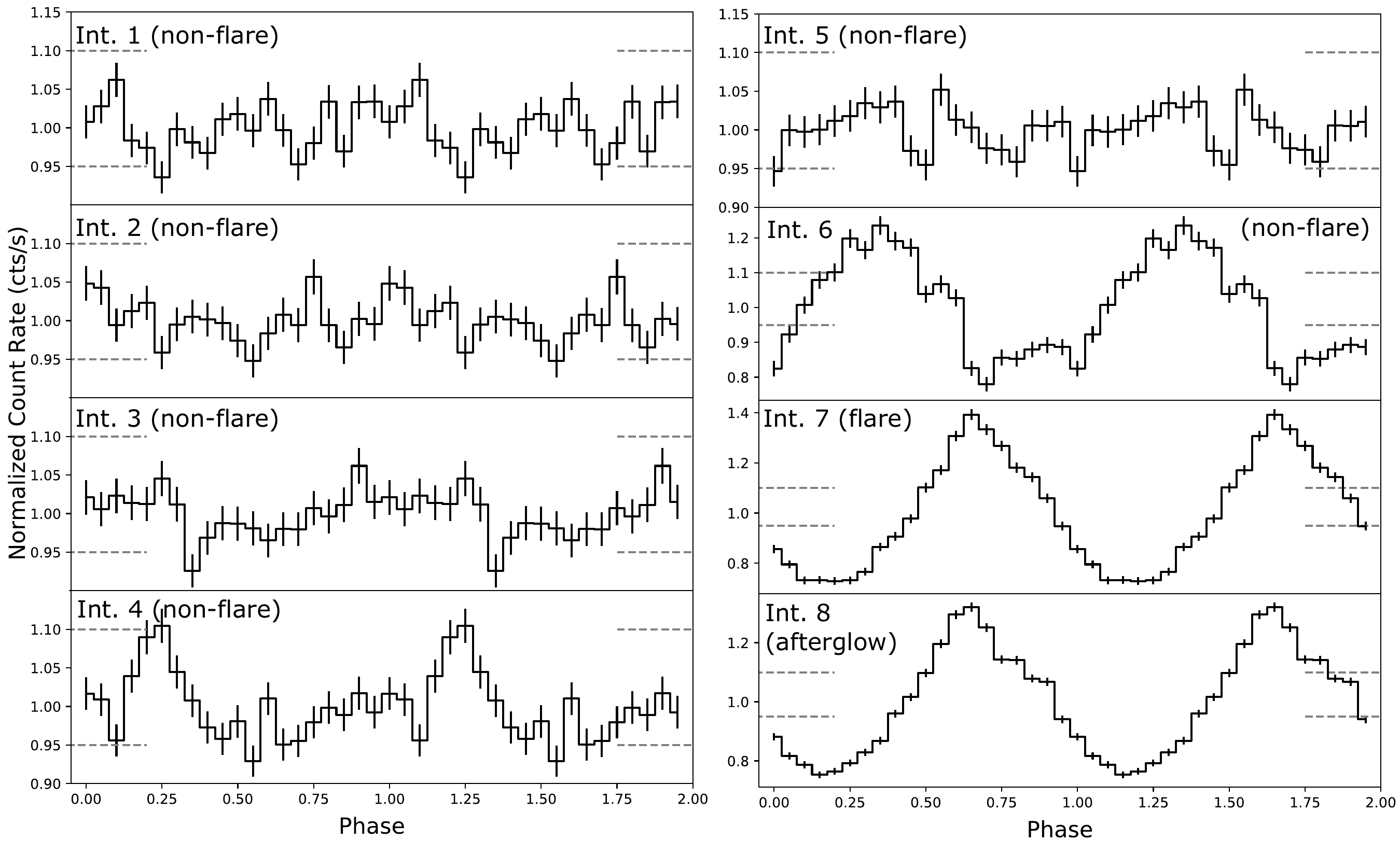}
\caption{Pulse profiles for Observation 08 in the full \nustar\ energy band, where the count rate has been normalized to the mean count rate in each interval. The horizontal dashed lines are plotted at the same rate in every panel to indicate the scale. The strength of the pulses dramatically increases in interval 6, and continues to increase with the onset of the flare. Changes in pulse shape are also evident, from more complex shapes in intervals 1--6 to a single pointy peak in intervals 7 and 8. A significant phase shift occurs between intervals 6 and 7.}
\label{fig:pulseprofs4}
\end{figure*}

We allowed the \nustar\ orbit to motivate the selection of time intervals over the course of each observation (see Figure \ref{fig:lchr}). Most of the intervals are the $\approx$4000 s intervals outside of Earth occultation, as selected by the pipeline. For times when the source flux changed rapidly, either due to the onset or cessation of a flare, we selected shorter time intervals based on flux. Where appropriate, we also divided the flare events into two time intervals that describe the main flare and the afterglow, as defined by \cite{shtykovsky2018}. The exception to this case is interval 2 in Observation 02, where the flare and the afterglow remain in the same interval due to the similarity in pulse behavior. We examined the pulse profile shape and strength in each of these intervals in order to asses the changes in pulse behavior over the course of these observations.

For every interval we determined the pulse profile by folding the interval's light curve by the spin period of the whole observation. The pulse profiles for each interval in the full \nustar\ energy band are shown in Figure \ref{fig:pulseprofs1} for Observation 02 and Figure \ref{fig:pulseprofs4} for Observation 08. We find a dramatic change in both shape and strength of the pulse profile across the observations. 

In Observation 02, the pulsations are strong in the first 7 intervals, during which the three accretion flares occur. The pulses are strongest and most pointy in the intervals that contain flares or afterglows: intervals 2, 4, 6, and 7. The intervals that immediately precede the flares (intervals 1, 3, and 5) also show strong pulsations, although the pulse shape is more variable. Following the third and final flare, the pulsations drop out completely for the rest of the observation (intervals 8--10). 

In Observation 08, intervals 1--5 show generally weak pulsations. In particular, intervals 1, 2, 3, and 5 do not have clear pulsations. However interval 6, the time immediately before the flare, shows pulsations approximately four times as strong as the pulses in the preceding intervals. The pulsations then further strengthen with the onset of the flare, and start to decrease in strength during the afterglow. Similar to Observation 02, the pulse shape during the flare and afterglow is quite pointy, while the pulses that precede in interval 6 have a more rounded shape.

In addition to changes in shape and intensity, the pulse profiles in both observations exhibit a phase shift of approximately 0.25 cycles with the onset of a flare. This can be seen most clearly in Observation 08, where the phase shift is evident between interval 6 and interval 7. However, the same phase shift is also visible at the onset of flares in Observation 02, between interval 1 and interval 2, and also between interval 5 and interval 6. We confirmed that this phase shift is not an artifact of our orbital timing correction algorithm by examining the barycentered data without the orbital time correction using the HENDRICS (\citealt{hendrics}) program {\fontfamily{qcr}\selectfont HENphaseogram}. The phase shift remains qualitatively unchanged even without the orbital timing correction. The phase shift and change in pulse strength were also independently found by \cite{shtykovsky2018}.

The phase shift seen in the pulse profiles is most likely tied to a rapid change in the spin period of the pulsar. To quantify the change in frequency, we used {\fontfamily{qcr}\selectfont HENphaseogram} to calculate the time of arrival of the pulsations in each interval. With these times of arrival, we used the software TEMPO2 (\citealt{edwards2006}) to refine the timing solution and calculate frequency derivatives and respective errors. In Observation 02, we found the absolute magnitude of the period derivatives to be $\approx3 \times 10^{-7}$ Hz s$^{-1}$ and $\approx3 \times 10^{-8}$ Hz s$^{-1}$ in Observations 02 and 08, respectively. These frequency derivatives are much larger than the long term derivative of $\sim1 \times 10^{-10}$ Hz s$^{-1}$ found by \cite{molkov2015}.

\subsection{Fractional RMS amplitude} \label{sec:pf}

\begin{figure*}
\plottwo{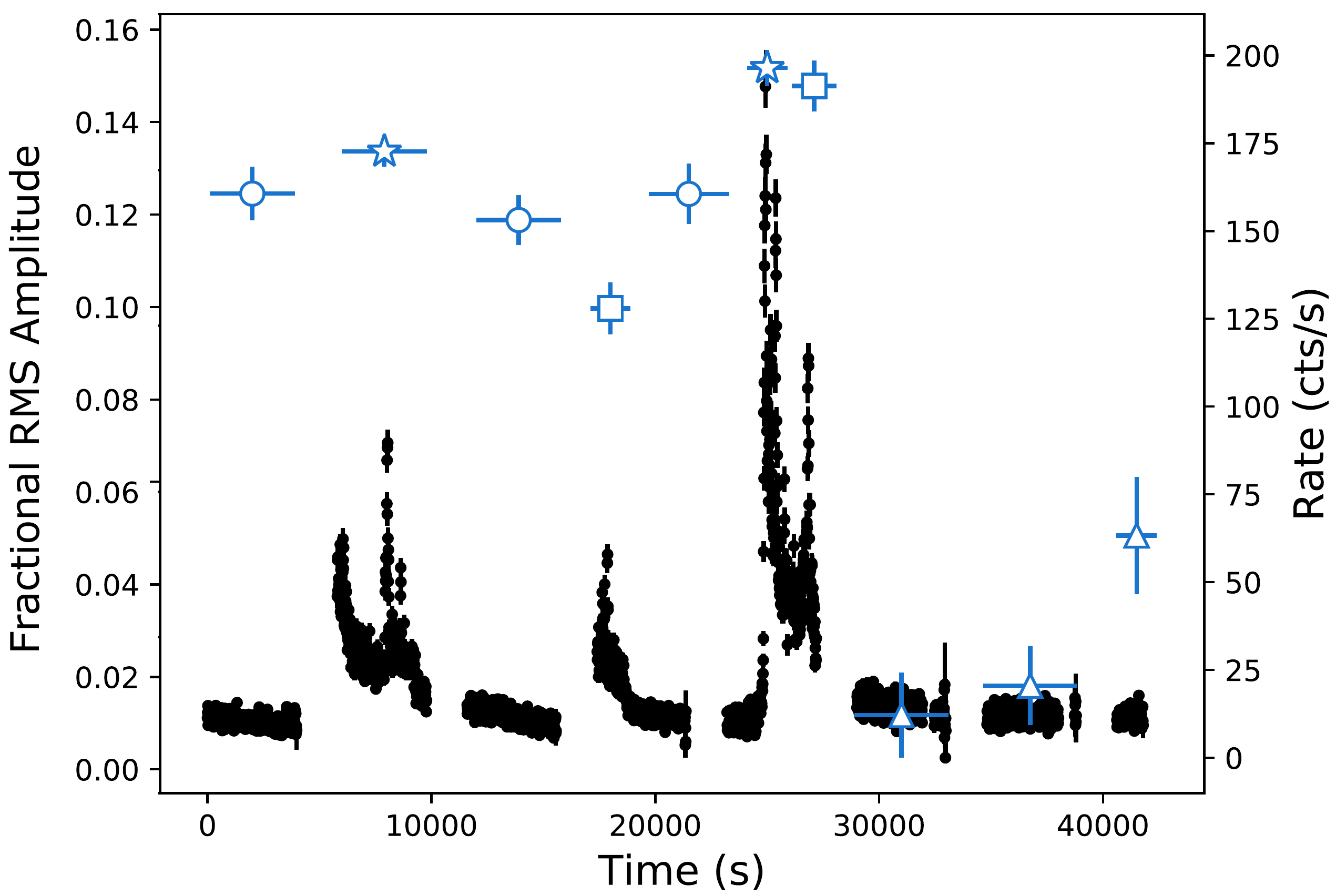}{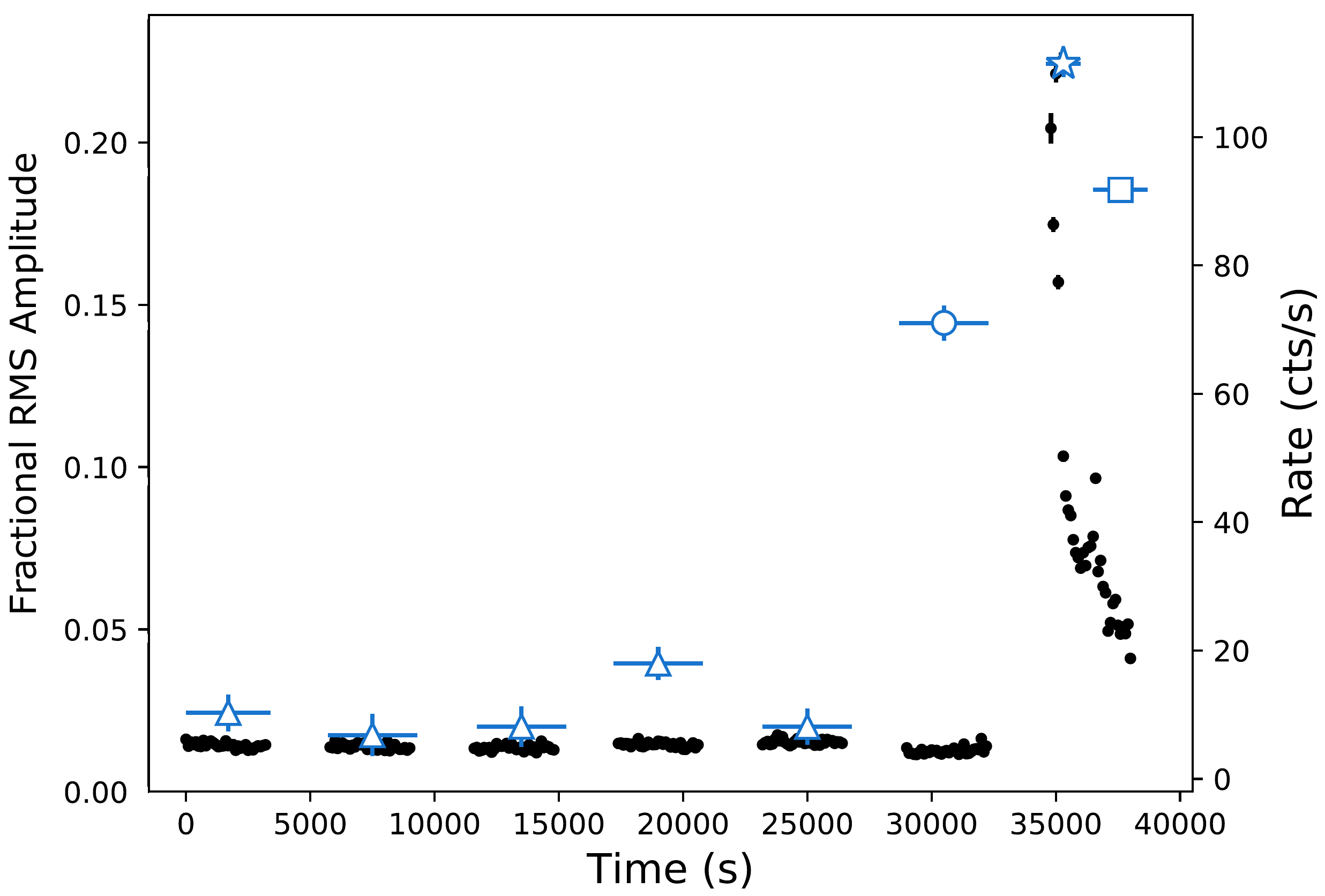}
\caption{Left: Fractional RMS amplitude vs. time for Observation 02. Right: Fractional RMS amplitude vs. time for Observation 08. For each figure, the corresponding light curve has been overplotted on the right axis for clarity. The fractional RMS amplitudes are shown in color for each time interval in the full \nustar\ energy range (left axis), where stars correspond to time intervals containing flares, squares to time intervals containing afterglows, circles to normal emission that is strongly pulsed, and triangles to normal emission that is weakly pulsed. The \nustar\ light curve for the observation is also plotted in black (right axis). The significant increase in pulsation strength immediately before and during the flare is evident.}
\label{fig:pfvst}
\end{figure*}

To quantify the changes in pulsations over the course of these observations, we computed the fractional RMS amplitude in each interval from the pulse profiles (e.g.\ \citealt{vaughan2003}). Figure \ref{fig:pfvst} shows that the fractional RMS amplitude is largest during the flares, reaching a maximum of 0.15 and 0.22 in Observations 02 and 08, respectively. The amplitude decreases in strength with the onset of the afterglow. The intervals immediately preceding flares (Observation 02: intervals 1, 3, and 5; Observation 08: interval 6) also show strong pulsations with a fractional RMS amplitude 0.12--0.15. However, following the third flare in Observation 02, the fractional RMS amplitude drops to below 0.05. We also find that the fractional RMS amplitude is below 0.05 in Observation 08 in intervals 1--5. The sudden dropout and turn on of pulsations in association with these flares can be seen in the pulse profiles as well (see Figures \ref{fig:pulseprofs1} and \ref{fig:pulseprofs4}).

To characterize the pulse behavior as a function of energy, we filtered the interval data into several energy bins: 3--5 keV, 5--7 keV, 7--10 keV, 10--15 keV, 15--20 keV, and 20--30 keV. In each of these energy bins, we used {\fontfamily{qcr}\selectfont efold} and our best spin period to generate pulse profiles for each interval. The weakly pulsed intervals in each observation (Observation 02: intervals 8--10; Observation 08: intervals 1--5) were combined due to the low significance of the pulsations. We then calculated the fractional RMS amplitude in the same way as before. We found very little energy dependence in the fractional RMS amplitude. Only the flare in Observation 08 showed a significant positive correlation with energy.

\section{Discussion} \label{sec:dis}
We observed several X-ray accretion flares across two \nustar\ observations of LMC X-4 during the turn-on of its superorbital high state. The flares show significant X-ray spectral softening and an increase in luminosity from $L_{X} \sim 1-2 \times 10^{38}$ erg s$^{-1}$ during normal activity to $L_{X} \sim 1 \times 10^{39}$ erg s$^{-1}$ during the flares. These characteristics are consistent with flares seen in previous studies (e.g.\ \citealt{levine2000}). While the origin of the flares is uncertain, \cite{moon2003} favor a scenario in which non-uniform stellar winds from the O star companion produce plasma Rayleigh-Taylor instabilities between the accretion disk and the magnetosphere, which ultimately give rise to a centrifugal barrier. The barrier allows material to accumulate in the accretion disk until a critical density is achieved, at which point accretion rapidly resumes and an X-ray burst is observed (\citealt{baan1977, baan1979}).

Our timing analysis indicates that pulsations dramatically drop out immediately following a large accretion flare in Observation 02 of LMC X-4. This phenomenon can be seen through the changes in shape and strength of the pulse profiles (Figure \ref{fig:pulseprofs1}) and the increase in fractional RMS amplitude (Figure \ref{fig:pfvst}). While the first seven intervals of Observation 02, where the flares occur, show strong pulsations, the fractional RMS amplitude drops below 0.05 in intervals 8--10. Intervals 8 and 9 do not seem to show coherent pulsations. We believe that we have observed a dropout of pulsations in association with the flare. This is the first known detection of pulse dropout in LMC X-4. 

The discovery of pulse dropout in Observation 02 is complimented by our analysis of Observation 08. In Observation 08 we find that the first five intervals show only weak pulsations, with fractional RMS amplitudes less than 0.05. This is similar to that seen in intervals 8--10 of Observation 02. This behavior changes in interval 6, where we find pulsations approximately four times as strong as those in previous intervals. The strength and shape of the pulsations in interval 6 is similar to those seen from the three pre-flare intervals of Observation 02 (intervals 1, 3, and 5). The flare itself and its afterglow are also strongly pulsed, as we saw in Observation 02. We believe that we have observed a ``turn-on" of pulsations for the first time in association with the flare.

In both observations, the transition from non- or weakly-pulsed to pulsed occurred during Earth occultation of the source. While this prevented us from observing the transition directly, it places an upper limit on the time scale of this transition of approximately 1500 s. We find it interesting that the intervals preceding flares in which pulsations are strong (Observation 02: intervals 1, 3, and 5; Observation 08: interval 6), in which the pulsations turn on, have an average count rate that is $\approx$20\% lower flux than the other pre-flare intervals. This slight decrease in source flux combined with an increase in pulse strength leads us to conclude that the propeller regime is not responsible for this pulse dropout, since this regime is characterized by pulsation dropout coinciding with a flux decrease. 

We also believe that obscuration is unlikely to be the cause of these pulse dropout events. Figure \ref{fig:lchr} indicates that the spectral shape is approximately constant across the last 3 time intervals of Observation 02 and the first 5 time intervals of Observation 08. Some spectral softening of approximately 20\% is seen in the pulsed, pre-flare intervals, although in Observation 4 the hardness ratio of interval 6 is only $\approx$10\% softer than interval 1. However, we expect that obscuration of the X-ray source would cause spectral hardening. Additionally, we do not believe that obscuration by the warped inner disk is involved because this instance of pulse dropout in LMC X-4 has occurred during the onset of the high state when the source is strongly detected, unlike Her X-1 where pulses are not detected in the low state due to obscuration by the inner accretion disk.

The pulse profiles of both Observation 02 and Observation 08 show a 0.25 cycle phase shift between intervals that contain flares and intervals that do not. Similar phase shifts have previously been seen during LMC X-4's flares by \cite{beri2017}. The phase shift visible in the pulse profiles could be caused by a number of effects. It has been previously noted that the hard and soft pulsations in LMC X-4 have an inherent phase offset due to the precession of the inner disk (\citealt{hung2010}). Because the flare is dominated by soft X-rays, it is possible that this leads to a phase shift during the flares. Another possible cause of the phase shift is changes in emission geometry of the accretion column. The super-Eddington luminosities seen in the flare events indicate that a significant increase in mass transfer rate has occurred, which could possibly affect the geometry of the emission in a way that would produce a phase shift in the pulsations. Models of accretion column emission that include relativistic light bending have found that changes in the emission geometry of the accretion column, or in the relative contribution of two accretion columns, can have a significant effect on pulse shape and phase in accreting X-ray pulsars (\citealt{falknersuba,falknersubb} submitted); \cite{dennerl1989} suggested that such effects could explain phase shifts in LMC X-4's pulse profiles during flares. These accretion column emission models have been found to well represent the energy-dependent pulse profiles of 4U 1626-67 (\citealt{iwakirisub}). 

Such changes in emission geometry could also feasibly be responsible for the apparent high period derivative during the flares. Magneto-hydrodynamic simulations of accreting neutron stars have shown that the location of the hotspot on the neutron star surface can change depending on the inner radius of the accretion disk, and that this movement can effect the observed spin frequency (\citealt{kulkarni2013}). This effect has been observed in accreting millisecond pulsars (e.g.\ \citealt{patruno2010}).

The complex pulse behavior seen from LMC X-4 in these observations raises many questions. It is not clear what mechanism caused the pulsations to drop out or turn on, however this behavior seems to be associated with the super-Eddington accretion flares. Further analysis of archival data is needed to more fully understand pulse dropout in LMC X-4 and its association with the X-ray flares.

The discovery of pulse dropout and a potential emission geometry change in LMC X-4 make it an excellent analog for studying the ultra-luminous X-ray (ULX) pulsar population. Because some of the brightest, non-ULX pulsars, such as LMC X-4 and SMC X-1, exhibit variable luminosities that can achieve super-Eddington rates, these sources are the best objects to bridge the gap between the ULX pulsars and standard accreting pulsars. Our detection of pulse dropout in association with super-Eddington X-ray flares suggests that a similar mechanism might be driving the pulse dropout in both LMC X-4 and the ULX pulsars. Further study of pulse dropout in LMC X-4 and the ULX pulsar population is needed to determine the origin of this phenomenon.

\section{Summary} \label{sec:sum}

In this Letter we perform a timing analysis on two observations of the HMXB LMC X-4 made with \nustar. Several accretion flares occurred during these observations, during which the source reaches super-Eddington luminosities. We generate pulse profiles and calculate the fractional RMS amplitude across these observations and conclude that in Observation 30102041008 the pulsations prior to the flare are extremely weak, even to the point of being undetected. We find that the pulsations return suddenly in the last 4000 s immediately preceding the flare, and continue to strengthen with the onset of the flare. This behavior is consistent with Observation 30102041002, where pulsations are strong immediately before each flare, but drop out completely after the final flare. We conclude that this is the first observation of pulse drop out and ``turn on" in LMC X-4 and suggest that the cause of this phenomenon could be linked to the origin of the flare. 

\acknowledgements
We would like to thank the anonymous referee for comments that substantially improved this Letter. We would also like to thank the NuSTAR Galactic Binaries Science Team for comments and contributions. MCB acknowledges support from NASA grant numbers NNX15AV32G and NNX15AH79H. This research made use of NuSTARDAS, developed by ASDC (Italy) and Caltech (USA) and of ISIS functions (ISISscripts) provided by 
ECAP/Remeis observatory and MIT (http://www.sternwarte.uni-erlangen.de/isis/).

\bibliography{my_bib}

\begin{thebibliography}{}
\expandafter\ifx\csname natexlab\endcsname\relax\def\natexlab#1{#1}\fi

\bibitem[{{Baan}(1977)}]{baan1977}
{Baan}, W.~A. 1977, \apj, 214, 245

\bibitem[{{Baan}(1979)}]{baan1979}
---. 1979, \apj, 227, 987

\bibitem[{{Bachetti}(2015)}]{hendrics}
{Bachetti}, M. 2015, {MaLTPyNT: Quick look timing analysis for NuSTAR data},
  Astrophysics Source Code Library, ascl:1502.021

\bibitem[{{Bachetti} {et~al.}(2014){Bachetti}, {Harrison}, {Walton},
  {Grefenstette}, {Chakrabarty}, {F{\"u}rst}, {Barret}, {Beloborodov}, {Boggs},
  {Christensen}, {Craig}, {Fabian}, {Hailey}, {Hornschemeier}, {Kaspi},
  {Kulkarni}, {Maccarone}, {Miller}, {Rana}, {Stern}, {Tendulkar}, {Tomsick},
  {Webb}, \& {Zhang}}]{bachetti2014}
{Bachetti}, M., {Harrison}, F.~A., {Walton}, D.~J., {et~al.} 2014, \nat, 514,
  202

\bibitem[{{Beri} \& {Paul}(2017)}]{beri2017}
{Beri}, A., \& {Paul}, B. 2017, New Astronomy, 56, 94

\bibitem[{{Bozzo} {et~al.}(2008){Bozzo}, {Falanga}, \& {Stella}}]{bozzo2008}
{Bozzo}, E., {Falanga}, M., \& {Stella}, L. 2008, \apj, 683, 1031

\bibitem[{{Carpano} {et~al.}(2018){Carpano}, {Haberl}, {Maitra}, \&
  {Vasilopoulos}}]{carpano2018}
{Carpano}, S., {Haberl}, F., {Maitra}, C., \& {Vasilopoulos}, G. 2018, \mnras,
  476, L45

\bibitem[{{Dall'Osso} {et~al.}(2015){Dall'Osso}, {Perna}, \&
  {Stella}}]{dallosso2015}
{Dall'Osso}, S., {Perna}, R., \& {Stella}, L. 2015, \mnras, 449, 2144

\bibitem[{{Deeter} {et~al.}(1998){Deeter}, {Scott}, {Boynton}, {Miyamoto},
  {Kitamoto}, {Takahama}, \& {Nagase}}]{deeter1998}
{Deeter}, J.~E., {Scott}, D.~M., {Boynton}, P.~E., {et~al.} 1998, \apj, 502,
  802

\bibitem[{{Dennerl}(1989)}]{dennerl1989}
{Dennerl}, K. 1989, in ESA Special Publication, Vol. 296, Two Topics in X-Ray
  Astronomy, Volume 1: X Ray Binaries. Volume 2: AGN and the X Ray Background,
  ed. J.~{Hunt} \& B.~{Battrick}

\bibitem[{{Edwards} {et~al.}(2006){Edwards}, {Hobbs}, \&
  {Manchester}}]{edwards2006}
{Edwards}, R.~T., {Hobbs}, G.~B., \& {Manchester}, R.~N. 2006, \mnras, 372,
  1549

\bibitem[{{Ek{\c s}i} {et~al.}(2015){Ek{\c s}i}, {Anda{\c c}}, {{\c
  C}{\i}k{\i}nto{\u g}lu}, {Gen{\c c}ali}, {G{\"u}ng{\"o}r}, \&
  {{\"O}ztekin}}]{eksi2015}
{Ek{\c s}i}, K.~Y., {Anda{\c c}}, {\.I}.~C., {{\c C}{\i}k{\i}nto{\u g}lu}, S.,
  {et~al.} 2015, \mnras, 448, L40

\bibitem[{{Falanga} {et~al.}(2015){Falanga}, {Bozzo}, {Lutovinov},
  {Bonnet-Bidaud}, {Fetisova}, \& {Puls}}]{falanga2015}
{Falanga}, M., {Bozzo}, E., {Lutovinov}, A., {et~al.} 2015, \aap, 577, A130

\bibitem[{{Falkner}(A submitted)}]{falknersuba}
{Falkner}, S. A submitted, \aap

\bibitem[{{Falkner}(B submitted)}]{falknersubb}
---. B submitted, \aap

\bibitem[{{F{\"u}rst} {et~al.}(2010){F{\"u}rst}, {Kreykenbohm}, {Pottschmidt},
  {Wilms}, {Hanke}, {Rothschild}, {Kretschmar}, {Schulz}, {Huenemoerder},
  {Klochkov}, \& {Staubert}}]{fuerst2010}
{F{\"u}rst}, F., {Kreykenbohm}, I., {Pottschmidt}, K., {et~al.} 2010, \aap,
  519, A37

\bibitem[{{F{\"u}rst} {et~al.}(2011){F{\"u}rst}, {Suchy}, {Kreykenbohm},
  {Barrag{\'a}n}, {Wilms}, {Pottschmidt}, {Caballero}, {Kretschmar},
  {Ferrigno}, \& {Rothschild}}]{fuerst2011}
{F{\"u}rst}, F., {Suchy}, S., {Kreykenbohm}, I., {et~al.} 2011, \aap, 535, A9

\bibitem[{{F{\"u}rst} {et~al.}(2016){F{\"u}rst}, {Walton}, {Harrison}, {Stern},
  {Barret}, {Brightman}, {Fabian}, {Grefenstette}, {Madsen}, {Middleton},
  {Miller}, {Pottschmidt}, {Ptak}, {Rana}, \& {Webb}}]{fuerst2016b}
{F{\"u}rst}, F., {Walton}, D.~J., {Harrison}, F.~A., {et~al.} 2016, \apjl, 831,
  L14

\bibitem[{{F{\"u}rst} {et~al.}(2017){F{\"u}rst}, {Kretschmar}, {Kajava},
  {Alfonso-Garz{\'o}n}, {K{\"u}hnel}, {Sanchez-Fernandez}, {Blay},
  {Wilson-Hodge}, {Jenke}, {Kreykenbohm}, {Pottschmidt}, {Wilms}, \&
  {Rothschild}}]{fuerst2017}
{F{\"u}rst}, F., {Kretschmar}, P., {Kajava}, J.~J.~E., {et~al.} 2017, \aap,
  606, A89

\bibitem[{{Giacconi} {et~al.}(1973){Giacconi}, {Gursky}, {Kellogg}, {Levinson},
  {Schreier}, \& {Tananbaum}}]{giacconi1973}
{Giacconi}, R., {Gursky}, H., {Kellogg}, E., {et~al.} 1973, \apj, 184, 227

\bibitem[{{Harrison} {et~al.}(2013){Harrison}, {Craig}, {Christensen},
  {Hailey}, {Zhang}, {Boggs}, {Stern}, {Cook}, {Forster}, {Giommi},
  {Grefenstette}, {Kim}, {Kitaguchi}, {Koglin}, {Madsen}, {Mao}, {Miyasaka},
  {Mori}, {Perri}, {Pivovaroff}, {Puccetti}, {Rana}, {Westergaard}, {Willis},
  {Zoglauer}, {An}, {Bachetti}, {Barri{\`e}re}, {Bellm}, {Bhalerao},
  {Brejnholt}, {Fuerst}, {Liebe}, {Markwardt}, {Nynka}, {Vogel}, {Walton},
  {Wik}, {Alexander}, {Cominsky}, {Hornschemeier}, {Hornstrup}, {Kaspi},
  {Madejski}, {Matt}, {Molendi}, {Smith}, {Tomsick}, {Ajello}, {Ballantyne},
  {Balokovi{\'c}}, {Barret}, {Bauer}, {Blandford}, {Brandt}, {Brenneman},
  {Chiang}, {Chakrabarty}, {Chenevez}, {Comastri}, {Dufour}, {Elvis}, {Fabian},
  {Farrah}, {Fryer}, {Gotthelf}, {Grindlay}, {Helfand}, {Krivonos}, {Meier},
  {Miller}, {Natalucci}, {Ogle}, {Ofek}, {Ptak}, {Reynolds}, {Rigby},
  {Tagliaferri}, {Thorsett}, {Treister}, \& {Urry}}]{harrison2013}
{Harrison}, F.~A., {Craig}, W.~W., {Christensen}, F.~E., {et~al.} 2013, \apj,
  770, 103

\bibitem[{{Hung} {et~al.}(2010){Hung}, {Hickox}, {Boroson}, \&
  {Vrtilek}}]{hung2010}
{Hung}, L.-W., {Hickox}, R.~C., {Boroson}, B.~S., \& {Vrtilek}, S.~D. 2010,
  \apj, 720, 1202

\bibitem[{{Illarionov} \& {Sunyaev}(1975)}]{illarionov1975}
{Illarionov}, A.~F., \& {Sunyaev}, R.~A. 1975, \aap, 39, 185

\bibitem[{{Israel} {et~al.}(2017{\natexlab{a}}){Israel}, {Belfiore}, {Stella},
  {Esposito}, {Casella}, {De Luca}, {Marelli}, {Papitto}, {Perri}, {Puccetti},
  {Castillo}, {Salvetti}, {Tiengo}, {Zampieri}, {D'Agostino}, {Greiner},
  {Haberl}, {Novara}, {Salvaterra}, {Turolla}, {Watson}, {Wilms}, \&
  {Wolter}}]{israel2017a}
{Israel}, G.~L., {Belfiore}, A., {Stella}, L., {et~al.} 2017{\natexlab{a}},
  Science, 355, 817

\bibitem[{{Israel} {et~al.}(2017{\natexlab{b}}){Israel}, {Papitto}, {Esposito},
  {Stella}, {Zampieri}, {Belfiore}, {Rodr{\'{\i}}guez Castillo}, {De Luca},
  {Tiengo}, {Haberl}, {Greiner}, {Salvaterra}, {Sandrelli}, \&
  {Lisini}}]{israel2017b}
{Israel}, G.~L., {Papitto}, A., {Esposito}, P., {et~al.} 2017{\natexlab{b}},
  \mnras, 466, L48

\bibitem[{{Iwakiri}(submitted)}]{iwakirisub}
{Iwakiri}, W. submitted, \apj

\bibitem[{{Kelley} {et~al.}(1983){Kelley}, {Jernigan}, {Levine}, {Petro}, \&
  {Rappaport}}]{kelley1983}
{Kelley}, R.~L., {Jernigan}, J.~G., {Levine}, A., {Petro}, L.~D., \&
  {Rappaport}, S. 1983, \apj, 264, 568

\bibitem[{{Kreykenbohm} {et~al.}(2008){Kreykenbohm}, {Wilms}, {Kretschmar},
  {Torrej{\'o}n}, {Pottschmidt}, {Hanke}, {Santangelo}, {Ferrigno}, \&
  {Staubert}}]{kreykenbohm2008}
{Kreykenbohm}, I., {Wilms}, J., {Kretschmar}, P., {et~al.} 2008, \aap, 492, 511

\bibitem[{{Kulkarni} \& {Romanova}(2013)}]{kulkarni2013}
{Kulkarni}, A.~K., \& {Romanova}, M.~M. 2013, \mnras, 433, 3048

\bibitem[{{Kuster} {et~al.}(2001){Kuster}, {Wilms}, {Staubert}, {Kreykenbohm},
  {Blum}, {Gruber}, \& {Rothschild}}]{kuster2001}
{Kuster}, M., {Wilms}, J., {Staubert}, R., {et~al.} 2001, in ESA Special
  Publication, Vol. 459, Exploring the Gamma-Ray Universe, ed. A.~{Gimenez},
  V.~{Reglero}, \& C.~{Winkler}, 309--312

\bibitem[{{Lang} {et~al.}(1981){Lang}, {Levine}, {Bautz}, {Hauskins}, {Howe},
  {Primini}, {Lewin}, {Baity}, {Knight}, {Rotschild}, \&
  {Petterson}}]{lang1981}
{Lang}, F.~L., {Levine}, A.~M., {Bautz}, M., {et~al.} 1981, \apjl, 246, L21

\bibitem[{{Levine} {et~al.}(1991){Levine}, {Rappaport}, {Putney}, {Corbet}, \&
  {Nagase}}]{levine1991}
{Levine}, A., {Rappaport}, S., {Putney}, A., {Corbet}, R., \& {Nagase}, F.
  1991, \apj, 381, 101

\bibitem[{{Levine} {et~al.}(2000){Levine}, {Rappaport}, \&
  {Zojcheski}}]{levine2000}
{Levine}, A.~M., {Rappaport}, S.~A., \& {Zojcheski}, G. 2000, \apj, 541, 194

\bibitem[{{Lutovinov} {et~al.}(2017){Lutovinov}, {Tsygankov}, {Krivonos},
  {Molkov}, \& {Poutanen}}]{lutovinov2017}
{Lutovinov}, A.~A., {Tsygankov}, S.~S., {Krivonos}, R.~A., {Molkov}, S.~V., \&
  {Poutanen}, J. 2017, \apj, 834, 209

\bibitem[{{Manousakis} \& {Walter}(2015)}]{manousakis2015}
{Manousakis}, A., \& {Walter}, R. 2015, \aap, 575, A58

\bibitem[{{Molkov} {et~al.}(2017){Molkov}, {Lutovinov}, {Falanga}, {Tsygankov},
  \& {Bozzo}}]{molkov2017}
{Molkov}, S., {Lutovinov}, A., {Falanga}, M., {Tsygankov}, S., \& {Bozzo}, E.
  2017, \mnras, 464, 2039

\bibitem[{{Molkov} {et~al.}(2015){Molkov}, {Lutovinov}, \&
  {Falanga}}]{molkov2015}
{Molkov}, S.~V., {Lutovinov}, A.~A., \& {Falanga}, M. 2015, Astronomy Letters,
  41, 562

\bibitem[{{Moon} \& {Eikenberry}(2001)}]{moon2001}
{Moon}, D.-S., \& {Eikenberry}, S.~S. 2001, \apjl, 549, L225

\bibitem[{{Moon} {et~al.}(2003){Moon}, {Eikenberry}, \& {Wasserman}}]{moon2003}
{Moon}, D.-S., {Eikenberry}, S.~S., \& {Wasserman}, I.~M. 2003, \apj, 586, 1280

\bibitem[{{Mushtukov} {et~al.}(2015){Mushtukov}, {Suleimanov}, {Tsygankov}, \&
  {Poutanen}}]{mushtukov2015}
{Mushtukov}, A.~A., {Suleimanov}, V.~F., {Tsygankov}, S.~S., \& {Poutanen}, J.
  2015, \mnras, 454, 2539

\bibitem[{{Nagase}(2001)}]{nagase2001}
{Nagase}, F. 2001, in American Institute of Physics Conference Series, Vol.
  556, Explosive Phenomena in Astrophysical Compact Objects, ed. H.-Y. {Chang},
  C.-H. {Lee}, M.~{Rho}, \& I.~{Yi}, 56--67

\bibitem[{{Patruno} {et~al.}(2010){Patruno}, {Hartman}, {Wijnands},
  {Chakrabarty}, \& {van der Klis}}]{patruno2010}
{Patruno}, A., {Hartman}, J.~M., {Wijnands}, R., {Chakrabarty}, D., \& {van der
  Klis}, M. 2010, \apj, 717, 1253

\bibitem[{{Quaintrell} {et~al.}(2003){Quaintrell}, {Norton}, {Ash}, {Roche},
  {Willems}, {Bedding}, {Baldry}, \& {Fender}}]{quaintrell2003}
{Quaintrell}, H., {Norton}, A.~J., {Ash}, T.~D.~C., {et~al.} 2003, \aap, 401,
  313

\bibitem[{{Ramsay} {et~al.}(2002){Ramsay}, {Zane}, {Jimenez-Garate}, {den
  Herder}, \& {Hailey}}]{ramsay2002}
{Ramsay}, G., {Zane}, S., {Jimenez-Garate}, M.~A., {den Herder}, J.-W., \&
  {Hailey}, C.~J. 2002, \mnras, 337, 1185

\bibitem[{{Romanova} {et~al.}(2004){Romanova}, {Ustyugova}, {Koldoba}, \&
  {Lovelace}}]{romanova2004}
{Romanova}, M.~M., {Ustyugova}, G.~V., {Koldoba}, A.~V., \& {Lovelace},
  R.~V.~E. 2004, \apj, 610, 920

\bibitem[{{Shtykovsky} {et~al.}(2018){Shtykovsky}, {Arefiev}, {Lutovinov}, \&
  {Molkov}}]{shtykovsky2018}
{Shtykovsky}, A.~E., {Arefiev}, V.~A., {Lutovinov}, A.~A., \& {Molkov}, S.~V.
  2018, Astronomy Letters, 44, 149

\bibitem[{{Staubert} {et~al.}(2004){Staubert}, {Kreykenbohm}, {Kretschmar},
  {Chernyakova}, {Pottschmidt}, {Benlloch-Garcia}, {Wilms}, {Santangelo},
  {Segreto}, {von Kienlin}, {Sidoli}, {Larsson}, \&
  {Westergaard}}]{staubert2004}
{Staubert}, R., {Kreykenbohm}, I., {Kretschmar}, P., {et~al.} 2004, in ESA
  Special Publication, Vol. 552, 5th INTEGRAL Workshop on the INTEGRAL
  Universe, ed. V.~{Schoenfelder}, G.~{Lichti}, \& C.~{Winkler}, 259

\bibitem[{{Tsygankov} {et~al.}(2016{\natexlab{a}}){Tsygankov}, {Lutovinov},
  {Doroshenko}, {Mushtukov}, {Suleimanov}, \& {Poutanen}}]{tsygankov2016b}
{Tsygankov}, S.~S., {Lutovinov}, A.~A., {Doroshenko}, V., {et~al.}
  2016{\natexlab{a}}, \aap, 593, A16

\bibitem[{{Tsygankov} {et~al.}(2016{\natexlab{b}}){Tsygankov}, {Mushtukov},
  {Suleimanov}, \& {Poutanen}}]{tsygankov2016a}
{Tsygankov}, S.~S., {Mushtukov}, A.~A., {Suleimanov}, V.~F., \& {Poutanen}, J.
  2016{\natexlab{b}}, \mnras, 457, 1101

\bibitem[{{Vaughan} {et~al.}(2003){Vaughan}, {Edelson}, {Warwick}, \&
  {Uttley}}]{vaughan2003}
{Vaughan}, S., {Edelson}, R., {Warwick}, R.~S., \& {Uttley}, P. 2003, \mnras,
  345, 1271

\bibitem[{{White}(1978)}]{white1978}
{White}, N.~E. 1978, \nat, 271, 38

\end{thebibliography}

\end{document}